# Performance Evaluation of Treecode Algorithm for N-Body Simulation Using GridRPC System

Truong Vinh Truong Duy, Katsuhiro Yamazaki, and Shigeru Oyanagi†

## 1. Introduction

GridSolve/NetSolve is a GridRPC supported middleware for parallel programming in a heterogeneous computing environment [3]. The purpose of GridSolve/NetSolve is to bring together disparate computational resources with a view to using their aggregate power and dominating the rich supply of services supported by the emerging Grid architecture. This paper is aimed at improving the performance of the treecode algorithm for N-Body simulation by employing the NetSolve GridRPC programming model to exploit the use of multiple clusters. N-Body is a classical problem, and appears in many areas of science and engineering, including astrophysics, molecular dynamics, and graphics. In the simulation of N-Body, the specific routine for calculating the forces on the bodies which accounts for upwards of 90% of the cycles in typical computations is eminently suitable for obtaining parallelism with GridRPC calls. It is divided among the compute nodes by simultaneously calling multiple GridRPC requests to them. The performance of the GridRPC implementation is then compared to that of MPI version and hybrid MPI-OpenMP version [1][2] for treecode algorithm on individual clusters. This paper presents the GridRPC system and then outlines parallelism achieving method for the treecode using GridRPC, MPI, and hybrid. Finally, it compares the experimental results and concludes our study.

## 2. The GridRPC computing system

The NetSolve middleware is utilized to construct a preliminary GridRPC computing system which consists of two existing clusters named Diplo and Raptor as depicted in Figure 1. The specifications of these two clusters are shown in Table I. The system comprises 2 agents, 12 servers with 24 processors. Basically, it is a RPC based client/agent/server system that allows users to remotely access both hardware and software components. At the top tier, the client library is linked in with the user's application which then makes calls to GridRPC API for specific services. Through the API, the client application gains access to aggregate resources without having to know how remote resources are involved. The tarbo and spino frontend nodes are designated as the primary and secondary agents, respectively. The agents maintain a database of all servers along with their capabilities and dynamic usage statistics which it uses to allocate server resources for client requests, ensuring load balancing and fault tolerance by keeping track of failed servers. The 4 compute nodes of Diplo and 8 compute nodes of Raptor operate as the servers of the system, executing remote functions on behalf of clients. A typical call from a client to the system involves several steps as illustrated in Figure 1.

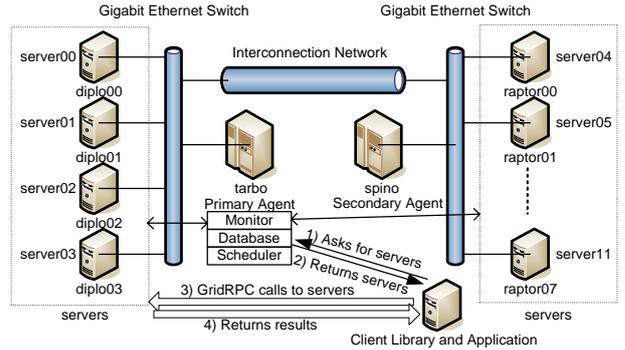

Fig. 1. The GridRPC computing system and a typical call from client.

## 3. The N-Body problem
### 3.1 Treecode algorithm

The n-body problem involves advancing the trajectories of n bodies according to their time evolving mutual gravitation field. The essence of the treecode [4] is the recognition that a distant group of bodies can be well-approximated by a single body, located at the center of mass with a mass equal to total mass of the group. It represents the distribution of the bodies in quad-tree for 2D space or oct-tree for 3D space. The tree is implemented by recursively dividing the 2D space into 4 subspaces, or 8 subspaces in 3D space, until the number of bodies in each subspace is below a certain threshold. Figure 2 demonstrates the distribution of bodies in 2D space and corresponding quad-tree.

After the tree construction phase, the force on a body in the system can be evaluated by traversing down the tree from root. At each level, a cell is added to an interaction list if the cell is distant enough for a force evaluation. Otherwise, the traversal continues recursively with the children. The accumulated list of interacting cells and bodies is then looped through to calculate

TABLE I: SYSTEM SPECIFICATIONS

| Name | SMP Node | # of Nodes | # of CPUs | Network |
|---|---|---|---|---|
| Diplo | Quad Xeon 3GHz | 4 | 16 | Gigabit Ethernet |
| Raptor | Pentium IV 3GHz | 8 | 8 | Gigabit Ethernet |

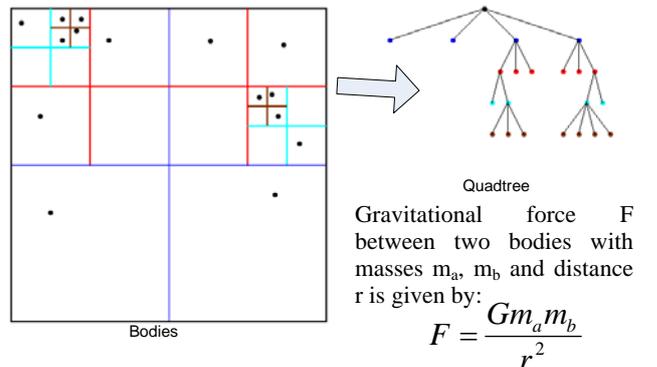

Gravitational force F between two bodies with masses $m_a$, $m_b$ and distance r is given by:

$$F = \frac{Gm_a m_b}{r^2}$$

Fig. 2. Bodies in 2D space and the quad-tree.

† Graduate School of Science and Engineering, Ritsumeikan University.

the force on the given body. Finally, each body updates its position and velocity based on the computed forces.

## 3.2 Exploiting parallelism with simultaneous GridRPC calls

The work load of force calculation which accounts for upwards of 90% of the cycles in typical computations is eminently suitable for obtaining parallelism with GridRPC calls and exploiting the use of multiple clusters. A simplified description of the GridRPC implementation is illustrated in Figure 3. After the tree construction phase for the original domain, the list of interactive nodes for each body is built by walking through the tree. Since the calculation of the summation force on each body with its list of interactive nodes is completely independent from each other, it is divided among the servers by simultaneously calling multiple asynchronous GridRPC requests to them. Actually parallelism achieving method in GridRPC system is quite similar to OpenMP parallelism in the hybrid MPI-OpenMP programming model as presented in the subsequent section, but utilizing simultaneous GridRPC calls over separate servers instead of parallel OpenMP threads on different processors.

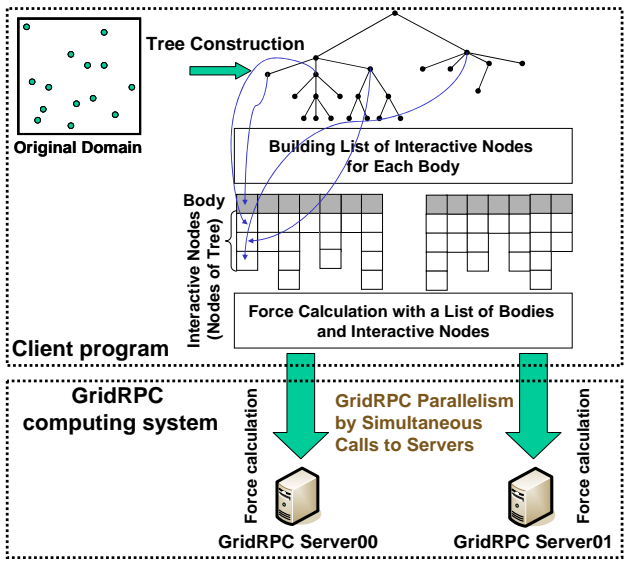

Fig. 3. Exploiting parallelism with simultaneous GridRPC calls.

In practice, there are two modules to be implemented, the server module for computing the forces and the client module for calling the server module. The server module receives input data, in particular the bodies and lists of interactive nodes, from the client module and returns the list of bodies with the updated force, position and velocity after each calculation. The implementation of the server module is deployed on all the servers together with the interface description file to build the library. The interface description file is then referred by the client program to learn how to correctly call the server module.

In the client module, the grpc_call_async API intended for asynchronous GridRPC calls is used in order to call the server module on servers remotely and concurrently. Consequently, the grpc_wait_all API is inserted after the asynchronous calls to wait for the completion of all of them for synchronization. Besides, since the GridSolve/NetSolve middleware integrates a self-scheduler running on the agents which automatically allocates remaining jobs to idle servers and keeps track of failed ones, it is unnecessary to explicitly implement the load balancing portion in the client module. The client program which assigns the bodies and corresponding lists of interactive nodes to execute on servers for force calculation using GridRPC calls is outlined as follows.

```
{
…
Constructs_the_tree(bodies);
Build_lists_of_interactive_nodes(bodies);
…
grpc_initialize();
grpc_function_handle_default("ComputeForces");
/* calling multiple asynchronous GridRPC requests
for force calculation on servers */
foreach body
 grpc_call_async(handle,current_body,current_list
 _of_interactive_nodes);
grpc_wait_all();
grpc_function_handle_destruct(handle);
grpc_finalize();
Move_bodies(bodies);
…
}
```

Figure 4 displays the time chart for the operation of the client program in conjunction with GridRPC calls to server module on the servers. Clearly, the initialization time and finalization time also contribute to the total computing time in addition to the computation time. However, they remain almost constant throughout the computation. Thus, the longer computation time is, the lower related overhead is, resulting in the higher parallelism achieved. This is one of the key aspects having an enormous impact on the performance of the GridRPC implementation.

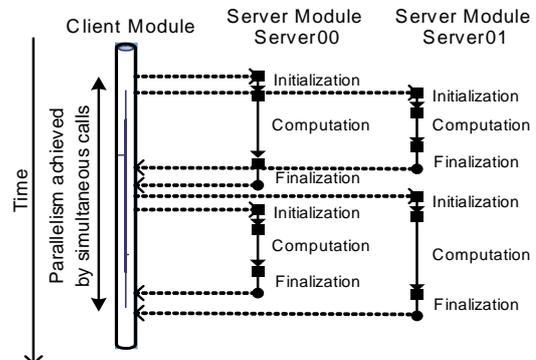

Fig. 4. Time chart of the client program with GridRPC parallelism.

### 3.3 MPI and hybrid versions for clusters

Prior to implementing the GridRPC N-Body simulation, we had developed the MPI program and hybrid MPI-OpenMP program for clusters [1][2]. In these cluster implementations, since the tree is very unbalanced when the bodies are not uniformly distributed in their bounding box, it is important to divide space into domains with equal work-loads to avoid load imbalance. Therefore, the Orthogonal Recursive Bisection (ORB) domain decomposition is adopted to divide the space into as many non-overlapping subspaces as processors, each of which contains an approximately equal number of bodies, and assign each subspace to a processor.

After the domain decomposition, each process has only the local tree for local bodies. In principle, they need the global tree to determine the forces due to the effect of influence ring along the borders. For example, node n belonging to process 0 has influence on bodies along the borders with P1, P2, and P3 as displayed in Figure 5. Thus node n, as well as other necessary nodes which represent a cluster of bodies, is called essential and must be known by P1, P2, and P3 to compute the forces of bodies in the influence ring of n. Each processor first collects all the nodes in its domain deemed essential to other processors by walking down its local tree from root, and then exchanges these essential nodes directly with the appropriate destination processors. Once all processors have received and inserted the data received into the local tree, each processor can proceed exactly as in the sequential case.

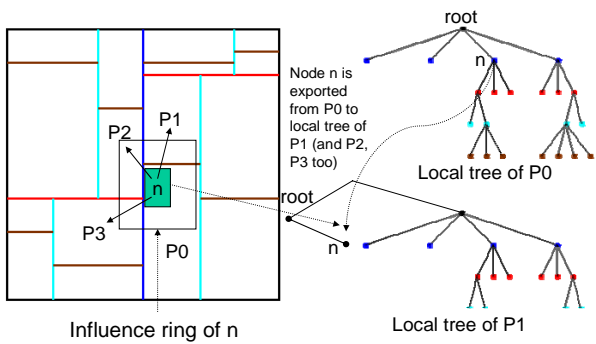

Fig. 5. ORB decomposition and the influence ring of a node.

Multiple levels of parallelism are achieved with the hybrid program as shown in Figure 6. For the first level, the bodies are distributed in a balanced way among the MPI processes using ORB domain decomposition. After the local trees have been constructed, the processes collect and exchange essential nodes to each other to insert into and expand the local trees. Each process then walks through its own tree to create a list of interactive nodes for each body similar to the case of sequential algorithm. For the second level, the force calculation is very appropriate for parallelizing with OpenMP work-sharing threads running in each MPI process. The bodies and their corresponding list of interactive nodes are assigned to different threads for calculating the force on each body.

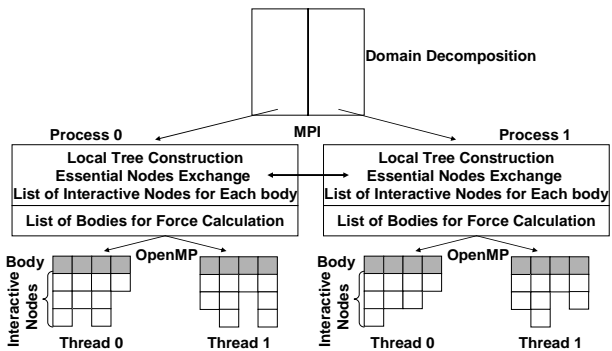

Fig. 6. Multiple levels of parallelism with the hybrid program.

## 4. Performance evaluation and discussion
### 4.1 Comparison between MPI and hybrid

The GridRPC N-Body implementation was executed on the GridRPC computing environment which is composed of the Diplo and Raptor clusters as presented in section 2 with the aim of evaluating and comparing the performance to the MPI and hybrid versions for clusters on Diplo. The timing runs of the GridRPC, MPI and hybrid MPI-OpenMP codes with the data sets of 10,000, 50,000, and 100,000 bodies for a fixed number (1, 2, 4, 8, 16, and 24) of CPUs are displayed in Figures 7, 8, and 9 respectively.

First of all, it is easy to recognize that no matter how many processors are used and how large the data set size is, the hybrid implementation outperforms the pure MPI one by an average of 30% faster on Diplo cluster at all times, stemming from the benefit of gaining two levels of parallelism, MPI level and OpenMP level. We also observed that performance of the hybrid program rises with the number of created OpenMP threads. Besides, another important advantage of the hybrid model compared to pure MPI model is that it lowers the number of sub-domains in ORB domain decomposition. For instance, we need to create only 4 sub-domains for the hybrid program while 16 sub-domains are necessary for the MPI program on 4-way Diplo cluster. As the number of sub-domains increases, the shapes of domains have a larger range of aspect ratios forcing tree walks to proceed to deeper levels. The complexity involved in determining locally essential data also rises with the number of sub-domains. We found that the number of node interactions grows with the number of sub-domains because of these effects. Thus, the hybrid model helps reduce this interaction overhead.

### 4.2 Comparison between GridRPC and other versions

Unlike the hybrid implementation which enjoys the benefit of multiple level parallelism regardless of data set sizes, the GridRPC program suffers a desperately poor performance with the execution time is almost twice as high as that of the hybrid program in case of the smallest data set of 10,000 bodies. The execution time is reduced gradually from 1 to 8 processors. However, this time rises when the number of processors is increased from 8 to 16 and from 16 to 24, leading to the best execution time achieved with 8 processors. The cause for that result is that the computation time for each GridRPC call on the servers is too short to take advantage of simultaneous calls and compensate the cost for initialization, communication and finalization, i.e., the parallel portion is not large enough compared to the sequential one of the program.

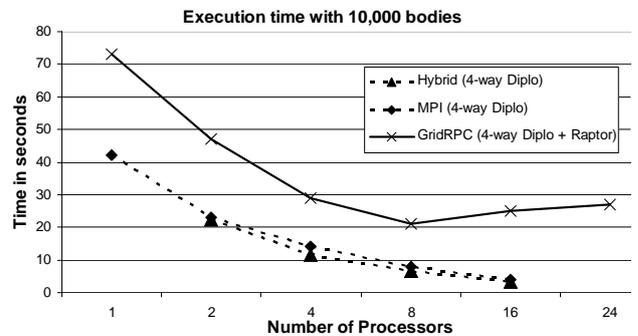

Fig. 7. Execution time of the GridRPC, MPI, and hybrid with 10,000 bodies.

When the data set grows to 50,000 bodies, more bodies require longer computation time for the GridRPC calls, causing an improved result of the GridRPC code. Even though the MPI and hybrid versions still strongly exhibit a better performance, the gap between them and the GridRPC implementation is narrowed. The GridRPC program is about 30% slower, but scales well at all times. The more processors are used, the lower execution time is. The peak performance is gained by 24 processors, giving rise to an execution time nearly equivalent to that of the MPI version with 16 processors.

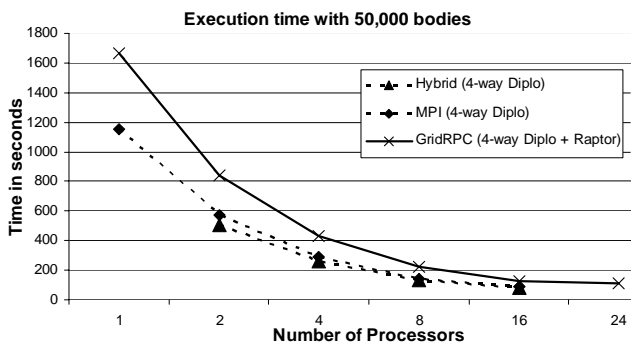

Fig. 8. Execution time of the GridRPC, MPI, and hybrid with 50,000 bodies.

The best performance of the GridRPC program is achieved with the largest data set of 100,000 bodies. Finally, this time it leaves both the MPI and hybrid versions behind with a significantly reduced execution time, approximately 40% and 10% faster, respectively. With the increase in the number of bodies, the parallel portion executed by the GridRPC servers now becomes large enough to effectively exploit the use of simultaneous calls. Meanwhile, the sequential time is trivial and does not contribute to the total computation time as much as in the previous data sets. Therefore, it is strongly expected that the GridRPC implementation will demonstrate even better performance with larger data sets. Besides, all the existing 24 processors of both clusters are employed to again successfully achieve the best result, a rise in the number of processors which can never be accomplished by individual clusters.

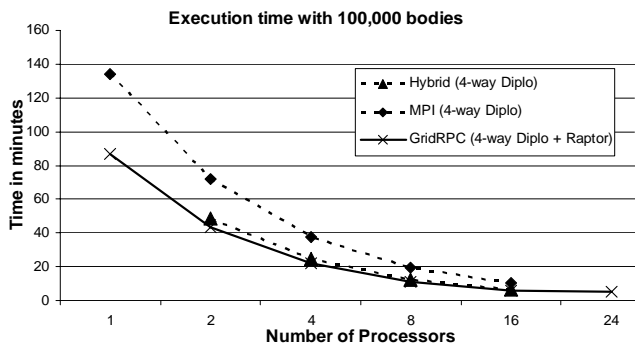

Fig. 9. Execution time of the GridRPC, MPI, and hybrid with 100,000 bodies.

### 4.3 Discussion

The performance evaluation and comparison show a quite interesting result. Given the ability to obtain multiple levels of parallelism, the hybrid program outperforms the corresponding pure MPI program whatever processors and data sets are used. Based on the fact that performance of the hybrid program rises with the number of created OpenMP threads, the difference factor between them is thought to be even higher in 8 or 16-way clusters although we have not had the opportunity to test the codes in such systems. On the other hand, we found that the size of the data set, that is, the computation time of the GridRPC calls executed on the servers, makes a big difference to the performance of the GridRPC implementation. In general, the GridRPC code scales well with an increase in the number of employed CPUs. However, the larger the data set is, the higher computation time is, bringing about the enhanced performance of the GridRPC program. With the largest data set of 100,000 bodies ever tested, the GridRPC performance becomes even better than that of MPI and hybrid codes. In addition, the lowest execution time is gained by an increased number of processors, 24, a great merit brought by the GridRPC model which outweighs the cluster model in terms of exploiting all the existing resources and producing higher throughput.

### 5. Conclusion

In this paper, we studied the programming efforts for N-Body simulation under the GridRPC programming paradigm and compared the performance among the GridRPC, MPI and hybrid MPI-OpenMPI implementations. In the GridRPC code, the specific routine for calculating the forces on the bodies is parallelized using multiple asynchronous calls on servers. We observed that the performance of GridRPC program is determined by the larger, the better size of data sets, in other words, the computation time for remote calls. The peak performance gained in case of 100,000 bodies is superior to the hybrid code's performance. The result once again proves that the GridRPC programming model is well suited to problems with intensively computing functions. Therefore, porting to a multiple cluster computing environment using GridRPC is a proper approach to maximize the use of resources. Also, a number of methods have been introduced in addition to treecode in which the FMM is shown to be only O(N). Adapting the parallel solutions using FMM is expected to speedup the performance.